\documentclass[a4paper,graphics,natbib,graphicx,psfig,amssymb,ccaption]{article}
\usepackage{lscape}
\usepackage{graphicx}
\newcommand{\affil}[1]{$^{\rm #1}$}
\usepackage[authoryear]{natbib}
\bibpunct{(}{)}{;}{a}{}{,}
\date{} 

\title{\large\bf\flushleft Metallicity Dependent Transformations for Red Giants with Synthetic Colours of {\em UBV} and {\em ugr}}
\author{\parbox{\textwidth}{\flushleft
\vspace{-0.5cm}
{\it S. Karaali\affil{\dag, A, B} and E. Yaz G\"ok\c ce\affil{A}}\\
\vspace{0.4cm}
{\small \affil{A}\,Istanbul University, Faculty of Sciences, Department of Astronomy and Space Sciences, 34119, Istanbul, Turkey}\\
{\small \affil{B}\,Email: karsa@istanbul.edu.tr}}}
\begin{document}
\twocolumn[
\begin{changemargin}{.8cm}{.5cm}
\begin{minipage}{.9\textwidth}
\vspace{-1cm}
\maketitle

\small{\bf Abstract:}
We present metallicity dependent transformation equations between UBV and SDSS 
$ugr$ colours for red giants with synthetic data. The ranges of the colours 
used for the transformations are 0.400 $\leq(B-V)_{0}\leq$ 1.460, 
-0.085$\leq(U - B)_{0}\leq$ 1.868, 0.291$\leq(g-r)_{0}\leq$ 1.326, and 
1.030$\leq(u - g)_{0}\leq$ 3.316 mag, and cover almost all the observational 
colours of red giants. We applied the transformation equations to six clusters 
with different metallicities and compared the resulting $(u -  g)_{0}$ colours 
with the ones estimated by the calibration of the fiducial sequences of the 
clusters. The mean and standard deviation of the residuals for all clusters are 
$<\Delta(u -  g)_0> =$-0.01 and $\sigma(u -  g)_{0} =$0.07 mag, respectively. 
We showed that interstellar reddening plays an important role on the derived colours. 
The transformations can be applied to clusters as well as to field stars. They can be 
used to extend the colour range of the red giants in the clusters which are 
restricted due to the saturation of SDSS data.  
     
\medskip{\bf Keywords:} stars: late-type - (stars:) giants - stars: general, techniques: photometric (Galaxy:) globular clusters: individual (M92, M5, M15, M71) - (Galaxy:) open clusters: individual (NGC 6791)
\medskip
\medskip
\end{minipage}
\end{changemargin}
]
\small
\let\thefootnote\relax\footnote{\small \affil{\dag}\,Retired.}
\section{Introduction}
 The Sloan Digital Sky Survey (SDSS) is one of the most widely used sky surveys. 
Also, it is the largest photometric and spectroscopic survey in optical wavelengths. 
Another widely used sky survey is the Two Micron All Sky Survey (2MASS; \cite{Skrutskie06}) 
which imaged the sky across infrared wavelengths. The third one which is astrometrically 
and photometrically important survey is {\em Hipparcos} \citep{Perryman97}, re-reduced recently by \cite{vanLeeuwen07}.    

SDSS obtains images almost simultaneously in five broad -  bands ($u$, $g$, $r$, $i$, and $z$) centred at 
3540, 4760, 6280, 7690, and 9250 \AA, respectively (\citealt{Fukugita06, Gunn98, Hogg01, Smith02}). 
The photometric pipeline \citep{Lupton01} detects the objects, matches the data 
from five filters and measures instrumental fluxes, positions and shape parameters. 
The magnitudes derived from fitting a point spread function (PSF) are accurate to 
about 2 per cent in $g$, $r$, and $i$, and 3 - 5 per cent in $u$ and $z$ for bright sources 
($<$ 20 mag). Data Release 5 (DR5) is almost 95 per cent complete for point sources 
to ($u$, $g$, $r$, $i$, $z$) =(22, 22.2, 22.2, 21.3, 20.5) mag (we remind to the reader 
the recent data release of SDSS, DR6-DR9). The median FWHM of the PSFs is about 
1.5 arcsec \citep{Abazajian04}. The data are saturated at about 14 mag in 
$g$, $r$, and $i$, and about 12 mag in $u$ and $z$ (see \cite{Chonis08}). 

The $ugriz$ passbands on which the main Sloan Digital Sky Survey (SDSS) with 2.5 m telescope 
is based are very similar, but not quite identical, to the $u'g'r'i'z'$ passbands with which 
the standard  Sloan photometric system was defined on the 1.0 m telescope of the USNO 
Flagstaff Station \citep{Smith02}. However, one can use the transformation equations 
in the literature to make necessary transformations between two systems (cf. \cite{Rider04}).

It has been customary to derive transformations between a newly defined photometric system 
and those that are more traditional, such as the Johnson -- Cousins' $UBVRI$ system. 
The first transformations derived between the SDSS $u'g'r'i'z'$ system and the 
Johnson -  Cousins' photometric system were based on the observations in $u'$, $g'$, $r'$, $i'$, 
and $z'$ filters \citep{Smith02}. An improved set of transformations between the observations 
obtained in $u'g'r'$ filters at the Isaac Newton Telescope (INT) at La Palma, Spain, and \cite{Landolt92} 
UBV standards is derived by \cite{Karaali05}. The INT filters were designed to reproduce 
the SDSS system. \cite{Karaali05} presented for the first time transformation equations 
depending on two colours. 

\cite{Rodgers06} considered two - colour or quadratic forms in their transformation 
equations. \cite{Jordi06} used SDSS DR4 and $BVRI$ photometry taken from different sources 
and derived population (and metallicity) dependent transformation equations between SDSS and $BVRI$ 
systems. \cite{Chonis08} used transformations from SDSS ugriz to $UBVRI$ not depending on 
luminosity class or metallicity to determine CCD zero - points. In \cite{Bilir08}, transformations 
between SDSS (and 2MASS) and $BVRI$ photometric systems for dwarfs are given. Finally, we refer the recent 
paper of \cite{Yaz10} where transformations between SDSS, 2MASS and $BVI$ photometric systems 
for late type giants are presented. 

Most of the transformations mentioned in the preceding paragraphs are devoted to dwarfs or 
the $UV$ -  band has not been considered in the case of giants. We thought to derive transformation 
equations between one of the most widely used sky surveys, SDSS $ugr$, and $UBV$ for giants. 
The saturation of the data in SDSS mentioned above restricts the range of the observed data in this system. 
Hence, we decided to use the synthetic $ugr$, as well as $UBV$ data. We used the procedure of \cite{Buser78} 
who derived two -  colour equations between $RGU$ and $UBV$ photometries. The sections are organized as follows: 
Data are presented in Section 2. Section 3 is devoted to the transformation equations and their application, 
and finally a summary and discussion is given in Section 4.               

\section{Data}
We used two sets of data. The $U - B$ and $B -  V$ synthetic colours are taken from \cite{Buser92}. 
\cite{Buser92} published the synthetic magnitudes and colours for 234 stars with different effective 
temperature, surface gravity and metallicity. The ranges of these parameters are 3750 $\leq T_{e} \leq$ 6000 K, 
0.75 $\leq$ log $g \leq$ 5.25 (cgs), and -3.00 $\leq \lbrack M/H\rbrack \leq$ 0.50 dex. We combined the $U -  B$, 
and $B -  V$, colours for the surface gravities log $g =$ 2.25 and 3.00 for the same temperature, 
and obtained a set of colours for the metallicities $\lbrack M/H\rbrack =$ 0.00, -1.00, and -2.00 
for six effective temperatures, i.e. 3750, 4000, 4500, 5000, 5500, and 6000 K for the red giants. 
This is the synthetic colour set in the $UBV$ system of our sample used in the transformations.

The $u -  g$ and $g -  r$ synthetic colours are provided from \cite{Lenz98}. The authors synthesized 
$u' -  g'$, $g' -  r'$, $r' -  i'$, and $i' -  z'$ colours with a large range of temperature, surface gravity, 
and metallicity using spectra from \cite{Kurucz91}. The range of the temperature is rather large, 
3500 $\leq T_{e} \leq$ 40 000 K, for the colours with surface gravities corresponding to dwarfs 
or red giants, i.e. log $g =$ 2.5, 3.0, 4.0, 4.5 (cgs), and metallicities $\lbrack M/H\rbrack =$ 
0.00, -1.00, -2.00 dex. However, a limited set of colours also exists for surface gravities log $g =$ 1.0, 1.5, 2.0 
(cgs) and metallicity $\lbrack M/H\rbrack =$ -5.00 dex. We followed the procedure explained 
in the preceding paragraph and combined the $u' - g'$, $g' -  r'$, and $r' - i'$ colours for the 
surface gravities log $g =$ 2.5 and 3.00 (cgs) for the same temperature, and obtained a set 
of colours for the metallicities $\lbrack M/H\rbrack =$ 0.00, -1.00, and -2.00 dex for six effective 
temperatures, i.e. 3750, 4000, 4500, 5000, 5500, and 6000 K for the red giants. This is the synthetic 
colour set in the $u'g'r'$ system of our sample. As mentioned in Section 1, the main SDSS with 2.5 m telescope is based on instrumental $ugriz$ passbands that are very similar, but not quite identical, to the $u'g'r'i'z'$ passbands with which the standard Sloan photometric system was defined on the 1.0 m telescope of the USNO Flagstaff Station \citep{Smith02}. Hence, we transformed 
the $u' - g'$ and $g' - r'$ colours of our sample to the $u - g$ and $g - r$ colours by the following 
equations derived by the equations in \cite{Rider04}:

\begin{eqnarray}
g - r = 1.060(g' - r') - 0.035 (r' - i') - 0.025,\nonumber\\
u - g = (u' - g') - 0.060(g' - r') + 0.032. 
\end{eqnarray}

The data and the corresponding two - colour diagrams in $UBV$ and $ugr$ systems are given in Table 1 and Fig. 1.  

\begin{table}[h]
\setlength{\tabcolsep}{4pt}
\center
\tiny{
 \caption{Synthetic colours as a function of effective temperature and metallicity.}
 \label{tabledata}

    \begin{tabular}{cccccccc}
    \hline
    $T_{e}$ & $U-B$ & $B-V$ & $u'-g'$ & $g'-r'$ & $r'-i'$ & $u-g$ & $g-r$ \\
    \hline
    \multicolumn{8}{c}{[M/H]=0 dex}                               \\
    \hline
    3750  & 1.868 & 1.460 & 3.357 & 1.228 & 0.572 & 3.316 & 1.256 \\
    4000  & 1.485 & 1.297 & 3.269 & 1.180 & 0.448 & 3.230 & 1.210 \\
    4500  & 0.892 & 1.033 & 2.629 & 0.892 & 0.306 & 2.607 & 0.910 \\
    5000  & 0.452 & 0.822 & 2.069 & 0.684 & 0.209 & 2.060 & 0.693 \\
    5500  & 0.212 & 0.655 & 1.634 & 0.517 & 0.142 & 1.635 & 0.518 \\
    6000  & 0.120 & 0.514 & 1.369 & 0.369 & 0.078 & 1.379 & 0.363 \\
    \hline
    \multicolumn{8}{c}{[M/H]=-1 dex}                              \\
    \hline
    3750  &  1.3530  & 1.356 & 3.061 & 1.293 & 0.549 & 3.015 & 1.326 \\
    4000  &  0.9820  & 1.180 & 2.706 & 1.093 & 0.450 & 2.673 & 1.117 \\
    4500  &  0.4750  & 0.916 & 2.049 & 0.788 & 0.317 & 2.034 & 0.799 \\
    5000  &  0.1670  & 0.724 & 1.366 & 0.606 & 0.228 & 1.362 & 0.610 \\
    5500  &  0.0190  & 0.568 & 1.256 & 0.454 & 0.157 & 1.260 & 0.450 \\
    6000  & -0.0115  & 0.435 & 1.121 & 0.318 & 0.091 & 1.134 & 0.309 \\
    \hline
    \multicolumn{8}{c}{[M/H]=-2 dex}                              \\
    \hline
    3750  &  1.176 & 1.339 & 2.770 & 1.242 & 0.562 & 2.727  & 1.272 \\
    4000  &  0.813 & 1.144 & 2.433 & 1.036 & 0.462 & 2.403  & 1.057 \\
    4500  &  0.326 & 0.879 & 1.794 & 0.766 & 0.333 & 1.780  & 0.775 \\
    5000  &  0.032 & 0.681 & 1.332 & 0.583 & 0.244 & 1.329  & 0.585 \\
    5500  & -0.087 & 0.522 & 1.094 & 0.427 & 0.165 & 1.100  & 0.422 \\
    6000  & -0.085 & 0.400 & 1.016 & 0.301 & 0.097 & 1.030  & 0.291 \\
    \hline
      \end{tabular}
}
 \label{tab:addlabel}
\end{table}

\begin{figure}
\begin{center}
\includegraphics[angle=0, width=70mm, height=117mm]{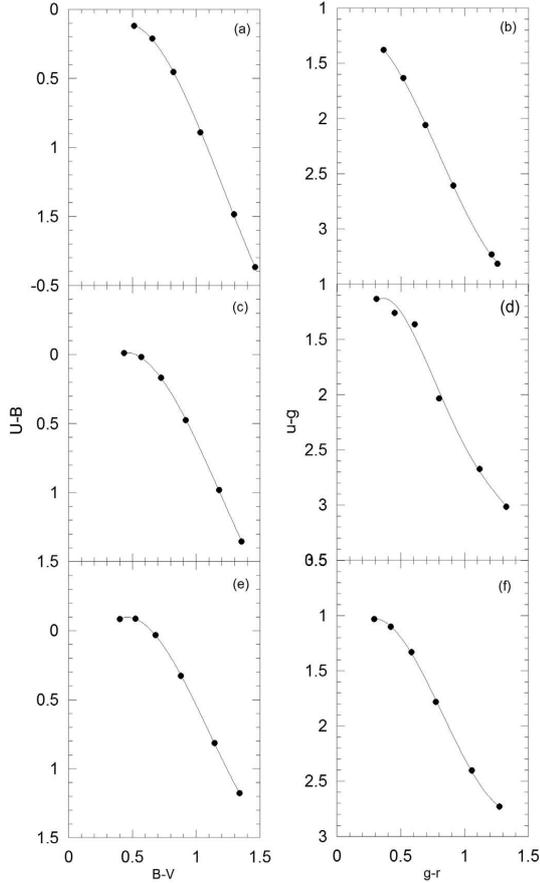}
\caption[] {$(U - B$, $B - V$) and ($u - g$, $g - r$) two - colour diagrams for different metallicities: $\lbrack M/H\rbrack =$ 0.00 (panels (a) and (b)), $\lbrack M/H\rbrack =$ -1.00 (panels (c) and (d)), $\lbrack M/H\rbrack =$ -2.00 (panels (e) and (f)).} 
\label{his:col}
\end{center}
\end {figure}

\section{Transformations}
\subsection{Transformation Equations from UBV to ugr}
We adopted the following general equations to transform the $U - B$ and $B - V$ colours 
to the $u - g$ and $g - r$ colours for three sets of data with different metallicities 
given in Table 1, i.e. $\lbrack M/H\rbrack =$ 0.00, -1.00, and -2.00 dex. 

\begin{eqnarray}
u - g = a ( U - B) + b ( B - V) + c,\nonumber\\ 
g - r = d ( U - B) + e ( B - V) + f.
\end{eqnarray} 

We used the least square method and evaluated the coefficients in Eq. (2) for each set of data. 
The results are given in Table 2. The ranges of $(U - B)$ and $(B - V)$ colours are also 
indicated in the  last two columns of the table. The numerical values of the coefficients 
a and b for the metallicities $\lbrack M/H\rbrack =$ -1.00 and -2.00 dex are close to 
each other indicating that both colours are effective in estimation of $u - g$ colour. 
However, for the metallicity $\lbrack M/H\rbrack =$ 0.00 dex, the value of b is almost 
9 times larger than the (absolute) value of a, which means that $B - V$ colour is much more effective 
than $U - B$ in estimation of colour $u - g$, for solar metallicities. The case is different 
in estimation of the colour $g - r$, i.e. the numerical value of e is at least 7 times larger
than the numerical value of $d$ for all metallicities. That is the colour $B - V$ plays 
much more role relative to the colour $U - B$ in estimation of the colour $g - r$. 
The mean of the residuals for $u - g$ and $g - r$ colours are almost zero, and the 
corresponding standard deviations are rather small, i.e. $\sigma(g-r)\leq$ 0.03 and 
$\sigma(u-g)\leq$ 0.09 mag.    

\begin{table*}[h]
\setlength{\tabcolsep}{4pt}
\center
 \caption{Numerical values for the coefficients in Eq. (2) for three different metallicities. The $U - B$ and $B - V$ intervals in the last two columns indicate the ranges of these colours.}
 \label{tabledata}
 \begin{tabular}{ccccccccc}
    \hline
    $\lbrack M/H\rbrack$ & $a$ & $b$ & $c$ & $d$ & $e$ & $f$ & $(U-B)$-range & $(B-V)$- range\\
    \hline
     0    & -0.329 & 2.820 & -0.070 & -0.187 & 1.342 & -0.312 & (0.120,1.868) & (0.514,1.460)\\
    -1    &  0.950 & 0.728 &  0.812 &  0.115 & 0.923 & -0.086 & (-0.012,1.353) & (0.435,1.356) \\
    -2    &  0.685 & 0.961 &  0.684 &  0.027 & 0.999 & -0.103 & (-0.085,1.176) & (0.400,1.339)\\
    \hline
    \end{tabular}
 \label{tab:addlabel}
\end{table*}

\subsection{Inverse Transformation Equations}

The general equations for the inverse transformations are adopted as follows:

\begin{eqnarray}
U - B = g (u - g) + h (g - r) + i,\nonumber\\ 
B - V = j (u - g) + k (g - r) + l.
\end{eqnarray} 

We applied the same procedure, i.e. the least square method, in evaluation of the numerical values 
of the coefficients in Eq. (3) for three different metallicities, $\lbrack M/H\rbrack =$ 0.00, -1.00, 
and -2.00 dex, in Table 1. The results are given in Table 3. The ranges of the colours $u - g$ and 
$g - r$ are also indicated in the last two columns of the table. Comparison of the corresponding 
coefficients shows that $g - r$ is effective in estimation of $U - B$ for metallicity $\lbrack M/H\rbrack =$ 
0.00 dex, while $u - g$ is more effective in estimation of the same colour for metallicity 
$\lbrack M/H\rbrack =$ -2.00 dex. Whereas both colours, $g - r$ and $u - g$, are equally effective 
for metallicity $\lbrack M/H\rbrack =$ -1.00 dex. The case is different in estimation of $B - V$, i.e.  
$g - r$ is much more effective relative to the colour $u - g$. The means of the residuals for $U - B$ 
and $B - V$ colours are almost zero and the standard deviations are small, i.e. $\sigma(U-B)\leq$ 0.06 
and $\sigma(B-V)\leq$ 0.04 mag, except the standard deviation for the colour $U - B$ for zero metallicity, 
i.e. $\sigma(U-B)=$ 0.15 mag.  

\begin{table*}[h]
\setlength{\tabcolsep}{4pt}
\center
 \caption{Numerical values for the coefficients in Eq.(3) for three different metallicities. The $u - g$ and $g - r$ intervals in the last two columns indicate the ranges of these colours.}
 \label{tabledata}
    \begin{tabular}{ccccccccc}
    \hline
  $\lbrack M/H\rbrack$ & $g$ & $h$ & $i$ & $j$ & $k$ & $l$ & $(u-g)$-range & $(g-r)$-range\\
    \hline
     0    & 0.283 &  1.279 & -0.889 & -0.164 & 1.369 & 0.222 & (1.379,3.316) & (0.363,1.256) \\
    -1    & 0.483 &  0.443 & -0,767 & -0.011 & 0.929 & 0.170 & (1.134,3.015) & (0.309,1.326) \\
    -2    & 1.009 & -0.513 & -1.005 &  0.016 & 0.934 & 0.114 & (1.030,2.727) & (0.291,1.272) \\
    \hline
      \end{tabular}
 \label{tab:addlabel}
\end{table*}

\subsection{Application of the Transformation Equations}
We applied the transformation equations to six clusters with different metallicities. 
The reason of preferring clusters instead of field stars is that they provide all the 
$U - B$, $B - V$, $u - g$, and $g - r$ colours necessary for transformations. 
Whereas, it is not easy to find a set of field stars with these colours and with 
different metallicities in the literature. The data for the clusters used in the 
application of the procedure is given in Table 4. The $U - B$ and $B - V$ colours 
are taken from the first reference, while the second reference (if it exists in the 
reference column) refers to the colour excess and metallicity.   

\begin{table}[h]
\setlength{\tabcolsep}{1pt}
\center
 \caption{Clusters used in the application of the procedure. The $U - B$ and $B - V$ colours are taken from the first reference, while the second reference (if it exists in the reference column) refers to the colour excess and metallicity.}
 \label{tabledata}
 \begin{tabular}{lccc}
    \hline
    Cluster & $E(B - V)$ & $[Fe/H]$ & Reference\\
    \hline
    M92     & 0.025 & -2.15 & (1), (2) \\
    M13     & 0.02  & -1.41 & (1), (2) \\
    M5      & 0.03  & -1.29 & (3)      \\
    M71     & 0.25  & -0.73 & (4), (5) \\
    M15     & 0.11  & -2.26 & (6)      \\
    NGC6791 & 0.13  &  0.37 & (3), (7) \\
    \hline
   \end{tabular}
\medskip\\
(1) \cite{Cathey74}, (2) \cite{Gratton97}, (3) \cite{vonBraun98}, (4) \cite{Hodder92}, (5) \cite{An08}, (6) \cite{Fahlman85}, (7) \cite{Sandage03}. \\
\label{tab:addlabel}
\end{table}

The transformation equations could be applied to the colours which 
fall into the ranges of the colours $(U - B)$ and $(B - V)$ stated 
in Table 2. The results are given in Table 5. The format for M13 is 
different than the other clusters due to the reason explained in the 
following. For evaluation of the $(u - g)_{0}$ and $(g - r)_{0}$ 
colours for the clusters M92, M15, M5, M71, and NGC 6791, we used 
the coefficients in Table 2 corresponding to the metallicities close 
to the metallicities of these clusters, i.e. $\lbrack M/H\rbrack =$ 
-2, -2, -1, -1, and 0 dex, while for the cluster M13 whose metallicity 
is in the middle of metallicity -1 and -2 dex, we used two sets of 
coefficients for each colour. One can notice small differences between 
the corresponding colours evaluated by means of two sets of coefficients. 

We evaluated the $(u - g)_{0}$ colours using another procedure and 
compared them with the ones  estimated by the transformation equations 
presented in this study, as explained in the following. We de - reddened 
the $(u - g)$ and $(g - r)$ colours in \cite{An08} for the clusters in Table 4 
and calibrated the $(u - g)_{0}$ colours in terms of $(g - r)_{0}$ ones for 
each cluster. Then, we applied these calibrations to the $(g - r)_{0}$ colours 
transformed from ($U -B)_{0}$ and $(B - V)_{0}$ colours. The $(u - g)_{0}$ 
colours thus obtained are labelled as $(u - g)_{0}(An)$. The residuals, i.e. 
the differences between the $(u - g)_{0}$ colours estimated by two different 
procedures, are given in the eighthly column for the clusters  M92, M15, 
M5, M71, and NGC 6791, while they are given in two columns, columns 8 and 12, 
for the cluster M13. Column 13 indicates the mean residuals (the other columns 
of Table 5 are explained at the top of this table). 

\begin{table*}[h]
\setlength{\tabcolsep}{1pt}
\center
\tiny{
 \caption{Transformation of the $U - B$ and $B - V$ colours to the $u - g$ and $g - r$ using the equations in (2). The columns give: (1) and (2) original $(B - V)$ and $(U - B)$ colours, (3) and (4) de - reddened $(B - V)_{0}$ and $(U - B)_{0}$ colours, (5) and (6) $(g - r)_{0}$ and $(u - g)_{0}$ colours estimated by the equations in (2), (7) $(u - g)_{0}$ colours evaluated by means of the calibrations of the fiducial sequences in \cite{An08}, and (8) the residuals, $\Delta(u-g)_{0}$. The procedure has been applied twice for the data of  M13, one for the metallicity $\lbrack M/H\rbrack =$-2 dex(columns (1) - (8) ) and one for $\lbrack M/H\rbrack =$-1 dex (columns (9) - (12)). Column (13) refers to the mean of the residuals evaluated for two cases.}
 \label{tabledata}
    \begin{tabular}{cccccccc|cccccccc}
    \hline
(1)   & (2)   & (3)   & (4)   & (5)   & (6)   & (7)   & (8)   & (1)   & (2)   & (3)   & (4)   & (5)   & (6)   & (7)   & (8) \\
    \hline
$B-V$& $U-B$&$(B-V)_{0}$&$(U-B)_{0}$&$(g-r)_{0}$&$(u-g)_{0}$&$(u-g)_{0}$(An)&$\Delta(u-g)_{0}$&$B-V$&$U-B$&$(B-V)_{0}$&$(U-B)_{0}$&$(g-r)_{0}$&$(u-g)_{0}$&$(u-g)_{0}$(An)&$\Delta(u-g)_{0}$\\
    \hline
    \multicolumn{8}{c|}{M92} & \multicolumn{8}{c}{M71}          \\
    \hline
    0.69  &  0.07  & 0.67  &  0.05 & 0.56 & 1.36  & 1.36 & 0.00 & 0.91 & 0.32 & 0.66 & 0.13 & 0.54 & 1.41 & 1.48 & -0.07 \\
    0.71  &  0.14  & 0.69  &  0.12 & 0.58 & 1.42  & 1.40 & 0.02 & 0.98 & 0.52 & 0.73 & 0.33 & 0.63 & 1.65 & 1.69 & -0.04 \\
    0.69  &  0.09  & 0.67  &  0.07 & 0.56 & 1.37  & 1.36 & 0.01 & 1.00 & 0.53 & 0.75 & 0.34 & 0.64 & 1.68 & 1.74 & -0.06 \\
    0.54  & -0.06  & 0.52  & -0.08 & 0.41 & 1.12  & 1.07 & 0.05 & 1.01 & 0.51 & 0.76 & 0.32 & 0.65 & 1.66 & 1.75 & -0.09 \\
    0.51  & -0.05  & 0.49  & -0.07 & 0.38 & 1.10  & 1.02 & 0.08 & 1.01 & 0.44 & 0.76 & 0.25 & 0.64 & 1.60 & 1.73 & -0.13 \\
\cline{1-8}    \multicolumn{8}{c|}{ M15}                        & 1.02 & 0.51 & 0.77 & 0.32 & 0.66 & 1.67 & 1.77 & -0.10 \\
\cline{1-8}0.8 & 0.13&0.69 & 0.04 & 0.59 &1.38&1.43 & -0.06 & 1.02 & 0.52  & 0.77 & 0.33  & 0.66  & 1.68  & 1.78  & -0.10 \\
    0.74  & 0.11  & 0.63  &  0.02 & 0.53 & 1.31  & 1.31  &  0.00 & 1.04 & 0.52 & 0.79 & 0.33 & 0.68 & 1.70 & 1.82  & -0.13 \\
    0.84  & 0.38  & 0.73  &  0.29 & 0.63 & 1.59  & 1.52  &  0.07 & 1.04 & 0.71 & 0.79 & 0.52 & 0.70 & 1.88 & 1.88  & 0.00 \\
    0.90  & 0.44  & 0.79  &  0.35 & 0.70 & 1.69  & 1.60  &  0.09 & 1.04 & 0.63 & 0.79 & 0.44 & 0.69 & 1.80 & 1.85  & -0.05 \\
    0.72  & 0.01  & 0.61  & -0.08 & 0.50 & 1.22  & 1.26  & -0.04 & 1.05 & 0.63 & 0.80 & 0.44 & 0.70 & 1.81 & 1.88  & -0.07 \\
    0.77  & 0.06  & 0.66  & -0.03 & 0.56 & 1.30  & 1.37  & -0.07 & 1.06 & 0.63 & 0.81 & 0.44 & 0.71 & 1.81 & 1.90  & -0.08 \\
\cline{1-8}    \multicolumn{8}{c|}{M5}                     & 1.07  & 0.63  & 0.82  & 0.44  & 0.72  & 1.82  & 1.92  & -0.10 \\
\cline{1-8} $-$ & $-$ & 0.56  & -0.066 & 0.423 & 1,157 & 1,157 & 0.000 & 1.08  & 0.65  & 0.83  & 0.46  & 0.73  & 1.85  & 1.95  & -0.10\\
  $-$ & $-$    & 0.74  & 0.264 & 0.627 & 1,602 & 1,583 & 0.018 & 1.09  & 0.73  & 0.84  & 0.54  & 0.75  & 1.93  & 2.00     & -0.07 \\
$-$ & $-$   & 0.74  & 0.289 & 0.63  & 1,625 & 1,590 & 0.035 & 1.12  & 0.81  & 0.87  & 0.62  & 0.79  & 2.03  & 2.09  & -0.06 \\
\cline{9-16}$-$ & $-$  & 0.75  & 0.244 & 0.634 & 1,590 & 1,599 & -0.009 & \multicolumn{8}{c}{NGC 6791}                                 \\
\cline{9-16}$-$ & $-$   & 0.78  & 0.264 & 0.664 & 1.631 & 1.666 & -0.036 &$-$&$-$ & 1.25  & 1.43  & 1.10 & 2.99 & 2.97 & 0.02\\
$-$ & $-$ & 0.855 & 0.469 & 0.757 & 1.880 & 1.854 & 0.026 & $-$ & $-$ & 1.26  & 1.49  & 1.09  & 2.98 & 2.96  & 0.02 \\
$-$ & $-$ & 0.910 & 0.504 & 0.812 & 1.953 & 1.939 & 0.014 & $-$ & $-$ & 1.26  & 1.40  & 1.12  & 3.02 & 3.00  & 0.02 \\
$-$ & $-$ & 0.915 & 0.529 & 0.819 & 1.981 & 1.949 & 0.032 & $-$ & $-$ & 1.32  & 1.66  & 1.14  & 3.09 & 3.04  & 0.05 \\
$-$ & $-$ & $-$ & $-$ & $-$ & $-$ & $-$ & $-$ & $-$ & $-$ &             1.34  & 1.61  & 1.18  & 3.17 & 3.10  & 0.07 \\
$-$ & $-$ & $-$ & $-$& $-$ & $-$ & $-$ & $-$ & $-$ & $-$ &              1.37  & 1.67  & 1.21  & 3.23 & 3.14  & 0.09 \\
$-$ & $-$ & $-$ & $-$ & $-$ & $-$ & $-$ & $-$ & $-$ & $-$ &            1.40  & 1.68  & 1.25  & 3.31 & 3.20  & 0.11 \\
\hline
    (1)   & (2)   & (3)   & (4)   & (5)   & (6)   & (7)   & \multicolumn{1}{c}{(8)} & (9)   & (10)  & (11)  & (12)  & (13) &  &  \\
\cline{1-13}$B-V$&$U-B$&$(B-V)_{0}$&$(U-B)_{0}$&$(g-r)_{0}$&$(u-g)_{0}$&$(u-g)_{0} (An)$&\multicolumn{1}{c}{$\Delta(u-g)$}& $(g-r)_{0}$ & $(u-g)_{0}$ & $(u-g)_{0}(An)$& $\Delta(u-g)$& $<\Delta(u-g)>$&  &  &  \\
\cline{1-13} \multicolumn{8}{c|}{(-2 dex)}& \multicolumn{4}{c|}{(-1 dex)}&  &   &  &  \\
\cline{1-13}    \multicolumn{13}{c}{M13}       &      &      & \\
\cline{1-13}0.75  & 0.23  & 0.73  & 0.21  & 0.63  & 1.53  & 1.58  & \multicolumn{1}{c}{-0.04} & 0.61  & 1.55  & 1.53  & 0.02 & -0.01 & & & \\
0.75&0.24&0.73&0.22&0.63&1.54& 1.58 & \multicolumn{1}{c}{-0.04} & 0.61  & 1.56  & 1.53  & 0.03  & -0.01 &  &  & \\
0.78&0.27&0.76&0.25&0.66&1.59& 1.66 & \multicolumn{1}{c}{-0.07} & 0.64  & 1.61  & 1.61  & 0.00  & -0.04 &  &  & \\
0.66&0.22&0.64&0.20&0.54&1.44& 1.35 & \multicolumn{1}{c}{0.09}  & 0.53  & 1.47  & 1.32  & 0.15  & 0.12  &  &  & \\
0.74&0.21&0.72&0.19&0.62&1.51& 1.55 & \multicolumn{1}{c}{-0.04} & 0.60  & 1.52  & 1.50  & 0.03  & -0.01 &  &  & \\
0.63&0.18&0.61&0.16&0.51&1.38& 1.29 & \multicolumn{1}{c}{0.10}  & 0.50  & 1.41  & 1.25  & 0.16  & 0.13  &  &  & \\
0.67&0.22&0.65&0.20&0.55&1.45& 1.38 & \multicolumn{1}{c}{0.07}  & 0.54  & 1.48  & 1.34  & 0.14  & 0.10  &  &  & \\
0.77&0.27&0.75&0.25&0.65&1.58& 1.63 & \multicolumn{1}{c}{-0.06} & 0.64  & 1.60  & 1.59  & 0.01  & -0.02 &  &  & \\
0.68&0.17&0.66&0.15&0.56&1.42& 1.40 & \multicolumn{1}{c}{0.03}  & 0.54  & 1.44  & 1.35  & 0.09  & 0.06  &  &  & \\
0.69&0.15&0.67&0.13&0.57&1.42& 1.42 & \multicolumn{1}{c}{0.00}  & 0.55  & 1.43  & 1.37  & 0.06  & 0.03  &  &  & \\
0.76&0.33&0.74&0.31&0.64&1.61& 1.61 & \multicolumn{1}{c}{0.00}  & 0.63  & 1.65  & 1.58  & 0.07  & 0.03  &  &  & \\
0.77&0.23&0.75&0.21&0.65&1.55& 1.63 & \multicolumn{1}{c}{-0.08} & 0.63  & 1.56  & 1.57  &-0.01  &-0.05  &  &  & \\
0.74&0.32&0.72&0.30&0.62&1.58& 1.56 & \multicolumn{1}{c}{0.03}  & 0.61  & 1.63  & 1.53  & 0.10  & 0.06  &  &  & \\
0.75&0.20&0.73&0.18&0.63&1.51& 1.57 & \multicolumn{1}{c}{-0.06} & 0.61  & 1.52  & 1.52  & 0.00  & -0.03 &  &  & \\
0.67&0.15&0.65&0.13&0.55&1.40& 1.37 & \multicolumn{1}{c}{0.03}  & 0.53  & 1.41  & 1.33  & 0.09  & 0.06  &  &  & \\
0.88&0.53&0.86&0.51&0.77&1.86& 1.99 & \multicolumn{1}{c}{-0.13} & 0.77  & 1.93  & 1.98  &-0.05  &-0.09  &  &  & \\
0.77&0.27&0.75&0.25&0.65&1.58& 1.63 & \multicolumn{1}{c}{-0.06} & 0.64  & 1.60  & 1.59  & 0.01  & -0.02 &  &  & \\
\hline
      \end{tabular}
}
 \label{tab:addlabel}
\end{table*}

The calibration of $(u - g)_{0}$ in terms of $(g - r)_{0}$ for the fiducial sequences of the clusters in questions is adopted as follows:

\begin{eqnarray}
(u - g)_{0} = m (g - r)_{0}^{3} + n( g - r)_{0}^{2}+ p(g - r)_{0} + q 
\end{eqnarray} 

The numerical values of the coefficients are given in  Table 6. The last column refers to the $(g - r)_{0}$ - range of the corresponding cluster. The $(g - r)_{0}$ colours estimated by  Eq. (2) which are beyond the range of the corresponding cluster could not be considered in our study and they are omitted from Table 5.    

The mean and the corresponding standard deviation of the residuals are $<\Delta(u - g)_{0} > =$ -0.01  and $\sigma( u - g)_{0} =$ 0.07 mag, respectively, i.e. the $(u - g)_{0}$ colour of a red giant would be estimated by the transformations presented in this study with an accuracy of $\Delta(u - g)_{0} <$ 0.1 mag.   

\begin{table}[h]
\setlength{\tabcolsep}{5pt}
\center
\tiny{
 \caption{Numerical values for the coefficients in Eq. (4).}
 \label{tabledata}
 \begin{tabular}{lcccccc}
    \hline
    Cluster & $m$ & $n$ & $p$ & $q$ & $R^{2}$    & $(g-r)_{0}$-int \\
    \hline
    M92      &   $-$   &    $-$   &  1.8693 &  0.3095 & 0.9991 & (0.37, 0.59) \\
    M13      &   $-$   &   2.2608 & -0.1745 &  0.7842 & 0.9992 & (0.34, 0.66) \\
    M5       & -7.2711 &  13.1300 & -5.6104 &  1.7308 & 0.9996 & (0.40, 0.88) \\
    M71      & -0.0815 &   0.6128 &  1.7352 &  0.3835 & 0.9991 & (0.38, 0.82) \\
    M15      & -12.282 &  18.6740 & -7.2682 &  1.7499 & 0.9990 & (0.42, 0.74) \\
    NGC 6791 &  3.4893 & -12.6550 & 16.8320 & -4.8743 & 0.9997 & (0.81, 1.29) \\
    \hline
      \end{tabular}
}
 \label{tab:addlabel}
\end{table}

\section{Summary and Discussion}
We presented metallicity dependent transformation equations from $UBV$ to $ugr$ 
colours and their inverse transformations for red giants with synthetic colours. 
The ranges of the colours used for the transformations are rather large, i.e. 
0.400 $\leq (B - V)_{0} \leq$ 1.460, -0.085 $\leq (U - B)_{0} \leq$ 1.868, 
0.291 $\leq (g - r)_{0} \leq$ 1.326, and 1.030 $\leq (u - g)_{0} \leq$ 3.316 mag, 
and cover almost all the  observational colours of red giants. Whereas, one can 
not obtain a set of observed colours with large range and with different metallicities 
available for transformations. We derived three sets of transformations for
 metallicities $\lbrack M/H\rbrack =$ 0, -1, and -2 dex. The researcher can use 
the transformation coefficients in Table 2 (or Table 3 for inverse transformations) 
regarding the metallicity of the red giant in question. One can also use two 
transformation equations and make an interpolation between two sets of results 
according to the metallicity of the star, similar to our calculations for the 
red giants in cluster M13 (section 3.3). 

We applied the procedure to two clusters of different metallicities as two examples, 
i.e. we derived the $(g - r)_{0}$ and $(u - g)_{0}$ two - colour diagrams for M5 ( Table 7) 
and NGC 6791 (Table 8) by transformation of the $(U - B)_{0}$ and $(B - V)_{0}$ colours of 
these clusters and plotted them in Fig.2 and Fig. 3, respectively. The points corresponding 
to the data evaluated via the red giant sequences in \cite{An08}, i.e. 0.4 $\leq (g - r_{0} \leq$ 0.82 
and  1.16 $\leq (u - g)_{0} \leq$ 1.98 mag for M5, and  1.09 $\leq (g - r)_{0} \leq$ 1.25 and 
2.98 $\leq (u - g)_{0} \leq$ 3.31 mag for NGC 6791 overlap to the diagrams, confirming our argument.

Also, we plotted the observed Johnson colours versus observed and predicted SDSS colours 
for the clusters M5 and NGC 6791 to see the trends of two sets of data. We illustrated 
the data for M5 in Fig.4 in two panels. Panel (a) gives the variations of the observed 
$(u - g)_{0}$ data (symbol: $+$) and the predicted ones (symbol: $\circ$) relative 
to the observed Johnson colour $(U - B)_{0}$, while those for $(g - r)_{0}$ colour relative 
to $(B - V)_{0}$ are shown in Panel (b) with similar symbols. The data for NGC 6791 are 
plotted in Fig.5 in panels (a) and (b), similar to Fig.4. We should note that the observed 
$(g - r)_{0}$ and $(u - g)_{0}$ colours are restricted with their ranges, as stated in the 
foregoing paragraph.

\begin{table}[h]
\setlength{\tabcolsep}{2pt}
\center
 \caption{$(u - g)_{0} - (g - r)_{0}$ two - colour red giant sequence 
of M5 derived from the $(U - B)_{0}$ and $(B - V)_{0}$ colours of the same 
cluster. The last column refers to the $(u - g)_{0}$ colours evaluated 
by means of the red giant sequence in \cite{An08}. The figures with bold 
face correspond to the $(g - r)_{0}$ colours which lie beyond the range 
of the cluster and which are not considered.}
 \label{tabledata}
    \begin{tabular}{ccccc}
    \hline
    $(B-V)_{0}$ & $(U-B)_{0}$ & $(g-r)_{0}$ & $(u-g)_{0}$ & $(u-g)_{0}$ (An) \\
    \hline
    0.56  & -0.07 & 0.42  & 1.16  & 1.16 \\
    0.74  & 0.26  & 0.63  & 1.60  & 1.58 \\
    0.74  & 0.29  & 0.63  & 1.63  & 1.59 \\
    0.75  & 0.24  & 0.63  & 1.59  & 1.60 \\
    0.78  & 0.26  & 0.66  & 1.63  & 1.67 \\
    0.86  & 0.47  & 0.76  & 1.88  & 1.85 \\
    0.91  & 0.50  & 0.81  & 1.95  & 1.94 \\
    0.92  & 0.53  & 0.82  & 1.98  & 1.95 \\
    0.98  & 0.61  & 0.88  & 2.10  & \bf{2.01} \\
    0.98  & 0.66  & 0.89  & 2.15  & \bf{2.01} \\
    1.01  & 0.76  & 0.93  & 2.27  & \bf{2.02} \\
    1.17  & 1.06  & 1.12  & 2.67  & \bf{1.72} \\
    1.25  & 1.18  & 1.20  & 2.84  & \bf{1.32} \\
    1.32  & 1.31  & 1.28  & 3.02  & \bf{0.79} \\
    1.35  & 1.30  & 1.31  & 3.03  & \bf{0.57} \\
    \hline
      \end{tabular}
 \label{tab:addlabel}
\end{table}

\begin{figure}
\begin{center}
\includegraphics[angle=0, width=50mm, height=70mm]{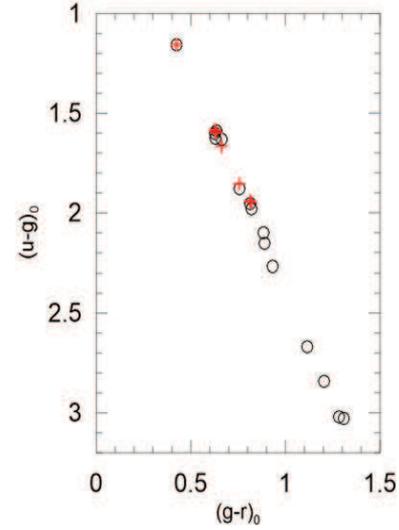}
\caption[]{The $(u - g)_{0} - (g - r)_{0}$ two - colour diagram of M5 based on the transformations in this study(symbol $\circ$). The points corresponding to the data evaluated via the red giant sequence in \cite{An08} are also plotted in this diagram (symbol $+$).} 
\label{his:col}
\end{center}
\end {figure}
%
\begin{figure}
\begin{center}
\includegraphics[angle=0, width=50mm, height=70mm]{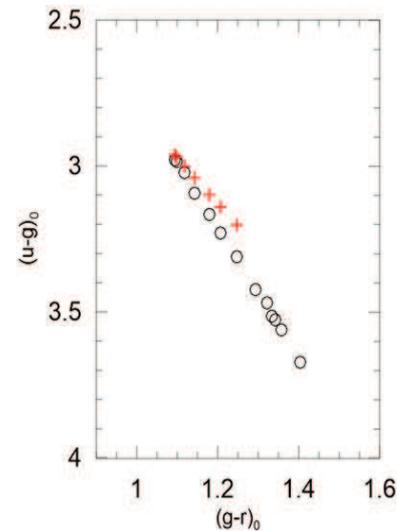}
\caption[]{The $(u - g)_{0} - (g - r)_{0}$ two - colour diagram of NGC 6791 based on transformations in this study (symbols as in Fig. 2).} 
\label{his:col}
\end{center}
\end {figure}
%
\begin{figure}
\begin{center}
\includegraphics[angle=0, width=90mm, height=60mm]{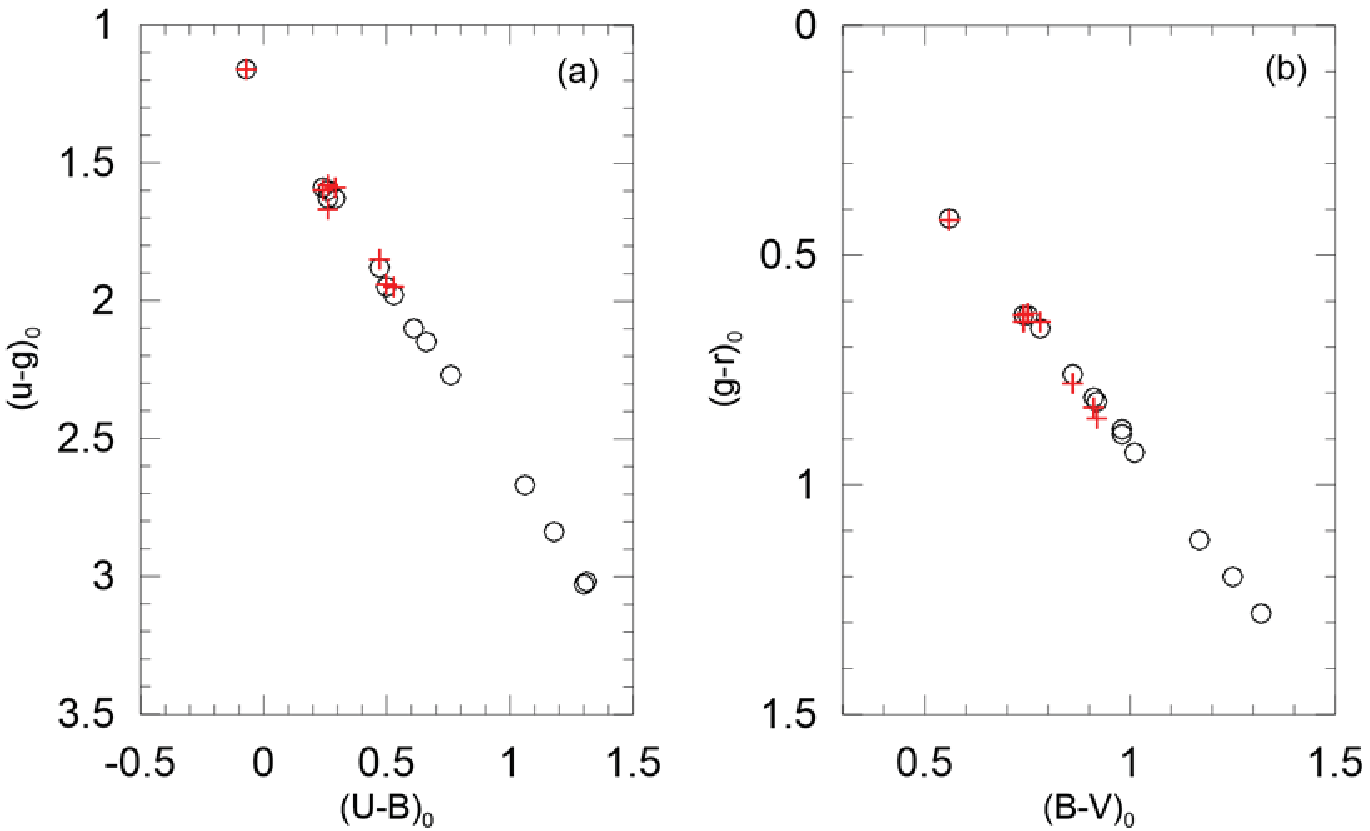}
\caption[]{Observed Johnson colours versus observed (symbol: $+$) and predicted (symbol: $\circ$) SDSS colours for cluster M5 in two panels: (a) $(U - B)_{0}$ versus $(u - g)_{0}$ , and (b) $(B - V)_{0}$ versus $(g - r)_{0}$.} 
\label{his:col}
\end{center}
\end {figure}
%
\begin{figure}
\begin{center}
\includegraphics[angle=0, width=90mm, height=60mm]{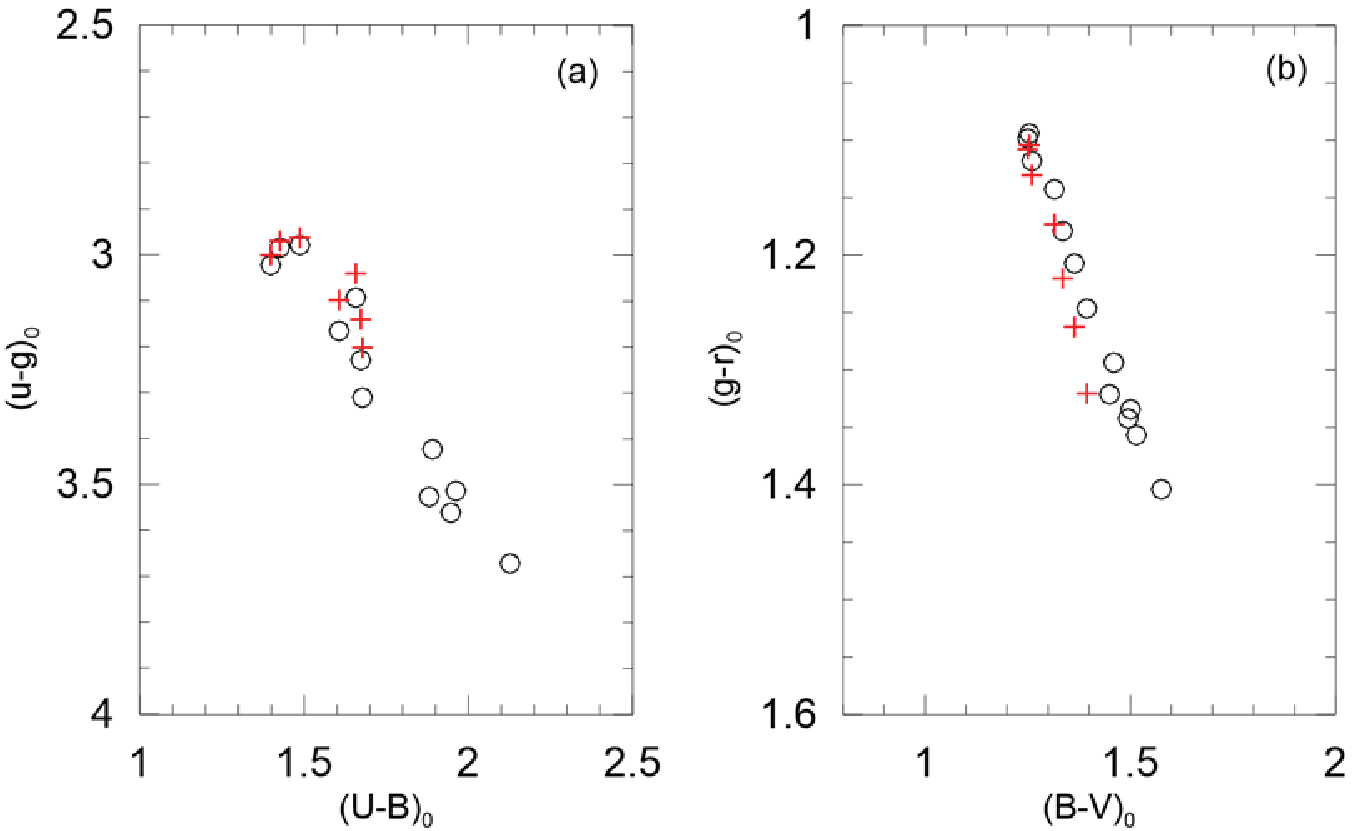}
\caption[]{Observed Johnson colours versus observed (symbol: $+$) and predicted (symbol: $\circ$) SDSS colours for cluster NGC 6791 in two panels: (a) $(U - B)_{0}$ versus $(u - g)_{0}$, and (b) $(B - V)_{0}$ versus $(g - r)_{0}$.}
\label{his:col}
\end{center}
\end {figure}
%

\begin{table}[h]
\setlength{\tabcolsep}{2pt}
\center
 \caption{$(u - g)_{0} - (g - r)_{0}$ two - colour red giant sequence of NGC 6791 derived from the (U - B)0 and (B - V)0 colours of the same cluster (the explanation of the columns are the same as in Table 7).}
 \label{tabledata}
\begin{tabular}{ccccc}
\hline
  $(B-V)_{0}$ & $(U-B)_{0}$ & $(g-r)_{0}$ & $(u-g)_{0}$ & $(u-g)_{0}$ (An) \\
\hline
    1.25  & 1.43  & 1.10  & 2.99  & 2.97 \\
    1.26  & 1.49  & 1.09  & 2.98  & 2.96 \\
    1.26  & 1.40  & 1.12  & 3.02  & 3.00 \\
    1.32  & 1.66  & 1.14  & 3.09  & 3.04 \\
    1.34  & 1.61  & 1.18  & 3.17  & 3.10 \\
    1.37  & 1.67  & 1.21  & 3.23  & 3.14 \\
    1.40  & 1.68  & 1.25  & 3.31  & 3.20 \\
    1.45  & 1.67  & 1.32  & 3.47  & \bf{3.32} \\
    1.46  & 1.89  & 1.29  & 3.42  & \bf{3.28} \\
    1.50  & 1.88  & 1.34  & 3.53  & \bf{3.36} \\
    1.50  & 1.96  & 1.33  & 3.51  & \bf{3.34} \\
    1.52  & 1.95  & 1.36  & 3.56  & \bf{3.38} \\
    1.58  & 2.13  & 1.40  & 3.67  & \bf{3.47} \\
\hline
      \end{tabular}
 \label{tab:addlabel}
\end{table}

\begin{table}[h]
\setlength{\tabcolsep}{2pt}
\center
 \caption{The mean and standard deviation of the residuals for each cluster and for their combination.}
 \label{tabledata}
    \begin{tabular}{lcc}
    \hline
    Cluster & $<\Delta(u - g)_{0}>$ & $\sigma(u - g)_{0}$ \\
    \hline
    M92      	 & 0.03  & 0.03 \\
    M13      	 & 0.02  & 0.06 \\
    M5       	 & 0.01  & 0.02 \\
    M71      	 &-0.08  & 0.03 \\
    M15      	 & 0.00  & 0.07 \\
    NGC 6791 	 & 0.05  & 0.04 \\
    All clusters &-0.01  & 0.07 \\
    \hline
      \end{tabular}
 \label{tab:addlabel}
\end{table}

The mean and the standard deviation of the residuals for each cluster as well as 
for their combination are given in Table 9. The ranges of the mean and the standard 
deviation of the residuals for all clusters are -0.08 $\leq \Delta (u - g)_{0} \leq$ 0.05 
and 0.02 $\leq \sigma(u - g)_{0} \leq$ 0.07 mag, respectively. While their mean and 
standard deviation are $<\Delta(u - g)_{0}> =$ -0.01 and $\sigma (u - g)_{0} =$ 0.07 mag, 
respectively. That is, the $(u - g)_{0}$ colours would be estimated by our transformations 
with an accuracy of $\Delta(u - g)_{0}\leq$ 0.08 mag. 

The probable parameter which would affect the colours estimated by the transformations 
in this study is the interstellar reddening. We confirmed our argument by applying the 
transformations to the data of the cluster M71 for two colour excesses, i.e. $E(B - V) =$ 0.25 
and 0.32 mag. which are taken from \cite{An08}. The corresponding standard deviations 
are equal, $\sigma(u - g)_{0} =$ 0.034 mag, whereas the mean of the residuals are 
different, i.e.  $<\Delta(u - g)_{0}> =$ -0.078 and -0.103 mag, for the colour excesses 
$E(B - V) =$ 0.25 and 0.32 mag, respectively. We adopted the colour excess $E(B-V) =$ 
0.25 mag in our statistics. 
                 
The transformation equations presented in this study can be applied to clusters 
as well as to field stars. It will be rather fruitful to derive the 
$(u - g)_{0} - (g - r)_{0}$ fiducial sequences of the red giants for some 
clusters whose metallicities are compatible with the metallicities of some populations. 
Cluster 47 Tuc can be given as an example for the Intermediate Population II, or thick disc. 
One can use the colour magnitude diagram of this cluster to evaluate the $M_{g}$ 
absolute magnitudes of the red giants of the thick disc population and estimate 
Galactic model parameters for this population.

\section*{Acknowledgments} 
We thank to anonymous referee for his/her comments and suggestions.

\end{document}